# High order dark wavefront sensing simulations

Roberto Ragazzoni[a,b], Carmelo Arcidiacono[b,c], Jacopo Farinato[a,b], Valentina Viotto[*,a,b], Maria Bergomi[a,b] Marco Dima[a,b], Demetrio Magrin[a,b], Luca Marafatto[a,b], Davide Greggio[a,b,d], Elena Carolo[a,b], Daniele Vassallo[a,b,d]

[a]INAF – Osservatorio Astronomico di Padova, Vicolo dell'Osservatorio 5, 35122, Padova, Italy;
[b]ADONI - Laboratorio Nazionale Ottiche Adattive, Italy;
[c]INAF - Osservatorio Astronomico di Bologna, Via Ranzani 1, I-40127 Bologna, Italy;
[d]Dipartimento di Fisica e Astronomia, Università degli Studi di Padova, Vicolo dell'Osservatorio 3, 35122, Padova, Italy;

## ABSTRACT

Dark wavefront sensing takes shape following quantum mechanics concepts in which one is able to "see" an object in one path of a two-arm interferometer using an as low as desired amount of light actually "hitting" the occulting object. A theoretical way to achieve such a goal, but in the realm of wavefront sensing, is represented by a combination of two unequal beams interferometer sharing the same incoming light, and whose difference in path length is continuously adjusted in order to show different signals for different signs of the incoming perturbation. Furthermore, in order to obtain this in white light, the path difference should be properly adjusted vs the wavelength used. While we incidentally describe how this could be achieved in a true optomechanical setup, we focus our attention to the simulation of a hypothetical "perfect" dark wavefront sensor of this kind in which white light compensation is accomplished in a perfect manner and the gain is selectable in a numerical fashion. Although this would represent a sort of idealized dark wavefront sensor that would probably be hard to match in the real glass and metal, it would also give a firm indication of the maximum achievable gain or, in other words, of the prize for achieving such device. Details of how the simulation code works and first numerical results are outlined along with the perspective for an in-depth analysis of the performances and its extension to more realistic situations, including various sources of additional noise.

## 1. INTRODUCTION

In WaveFront (WF) sensing the ultimate device is always a detector that actually converts photons optically processed by some optomechanical system into counts of events that can be associated with photons through a certain relationship. Ideally, a 100% Quantum Efficiency and the full absence of any further kind of noise (reading, dark, background, just to name some of the most known in the field) would still suffer from Poissonian noise, in the detection area. Furthermore, WF sensors, are just optomechanical devices that can split, select, fold, polarize, etc. incoming photons, but they do not have any ability to generate them, although one could identify some even more exotic kind of WF sensing using down-conversions or other non-linear techniques that are not exploited here. The way the Poissonian error would translate back into an associated error in the WF measurement would depend on upon how the selection of the photons to be detected is achieved.

We have already shown in the past that in some simple and crude WaveFront sensing[1][2], for instance in a four quadrant tip-tilt sensor, under certain conditions, most of the detected light is not giving any contribution to the actual signal, although it is contributing to the overall measurement noise under a Poissonian detection scheme. This means that these kind of detectors are –from a sort of fundamental viewpoint- working in an under-perfect regime and that their ultimate sensitivity is not yet achieved. We show elsewhere[3] that removing the flux of the photons that are not contributing to the signal can lead to an equivalent gain in sensitivity such that the same measurement noise can be achieved with an order of magnitude less light, or –using the astronomical jargon- with a 2.5 magnitude increase in the limiting magnitude.

---

[*] roberto.ragazzoni@oapd.inaf.it

This is a large achievement in terms of sky coverage when using Natural Guide Stars, and, whenever such techniques could be applicable to artificially generated Laser Guide Stars, they could translate into a tremendously beneficial effect in terms of laser power requirements.

Provided that there is still room for sensitivity improvement in WaveFront Sensors (WFSs hereafter), we introduced a new class of these devices, generically identified as Dark WFS (or DWFS hereafter) whose key feature is that when a flat wavefront is being analyzed, the actual photon detection map is close, or identical, to a map of zeros, no detections at all.

It is interesting to note a couple of important facts. The first is that this approach has been already investigated in quantum mechanical realms, leading to the concept of "seeing" objects without actually throwing photons at them[4]. Although this sounds particularly weird it is to be pointed out that this is strictly true only in a sort of "limit" sense, where the use of this word is used following the jargon used in mathematics for the concept of limit. In fact, "some" photons would actually have chances to hit the objects, but their amount can be made, in principle, arbitrarily low. The second is that the double arm or double Smartt possible implementations are just a way to show that a DWFS can be actually be built, but there is no indication that this, or variations on this theme, are the only actual way to do so. This second concern lead to a sort of semantic issue that I am not going either to precisely state nor to solve here, that essentially is that the double arm, or double Smartt[5][6] cannot be identified as "the" DWFS but just "one kind of DWFS". As this is the only known example of this class I would not propose, for the moment, any particular acronym or specific term. The development in the field will tell if the DWFS class is a rich and interesting one, or if it is populated just by few examples, destined to remain in the realm of the academic speculations rather than to rugged devices in use at some telescope's focal plane.

## 2. A DOUBLE ARM INTERFEROMETER

An ideal DWFS would show off two map's images onto which some light is popping out whenever the deviation from the flat wavefront is positive or negative. The amount of the deviation, in this case, would be given by the amount of the collected light in the portion of the pupil's images, while the sign would be dictated by which of the two maps is illuminated in the particular part. Although we didn't find a way to generate such processing of the light, we found that it is relatively easy to achieve something rather close to that. In practice the incoming light is divided into two arms and, in the following, it is assumed that the perturbation is smaller than a single wavelength. This translates into working at high Strehl Ratio, a situation that is becoming more and more common nowadays. The light is now entering into two arms, that we will name "positive" and "negative" arms although the difference will be subtle. The two arms are both interferometers in which one side of the further split light is made flat by passing through a spatial filtering. In Fig.1 this is accomplished by two optics that makes the light focusing onto a spatial filter that is supposed to be of the right size, with respect to the focal ratio of the lenses, in order to ensure that the outcoming wavefront, later used as reference, is basically flat (or at least much better than the deviation we are aiming to measure).

The interference, happening at the second beam-splitter of the two almost identical arms, is adjusted in a way that there is perfect destructive interference on one side and perfect constructive one when a flat wavefront is incoming. This condition is now perturbed by a quantity that in the following is called "Z shift" that is applied in conflicting directions in the two arms. If this Z-shift is, say, 50nm, a perturbation of such an amount would exhibit a perfect destructive interference in one arm, and would exhibit a total deviation of twice as much, in this case, 100nm, in the other. With a flat incoming wavefront, the destructive interference is not zero, so some light will go through, but this can be done as small as possible making the Z-shift going toward zero.

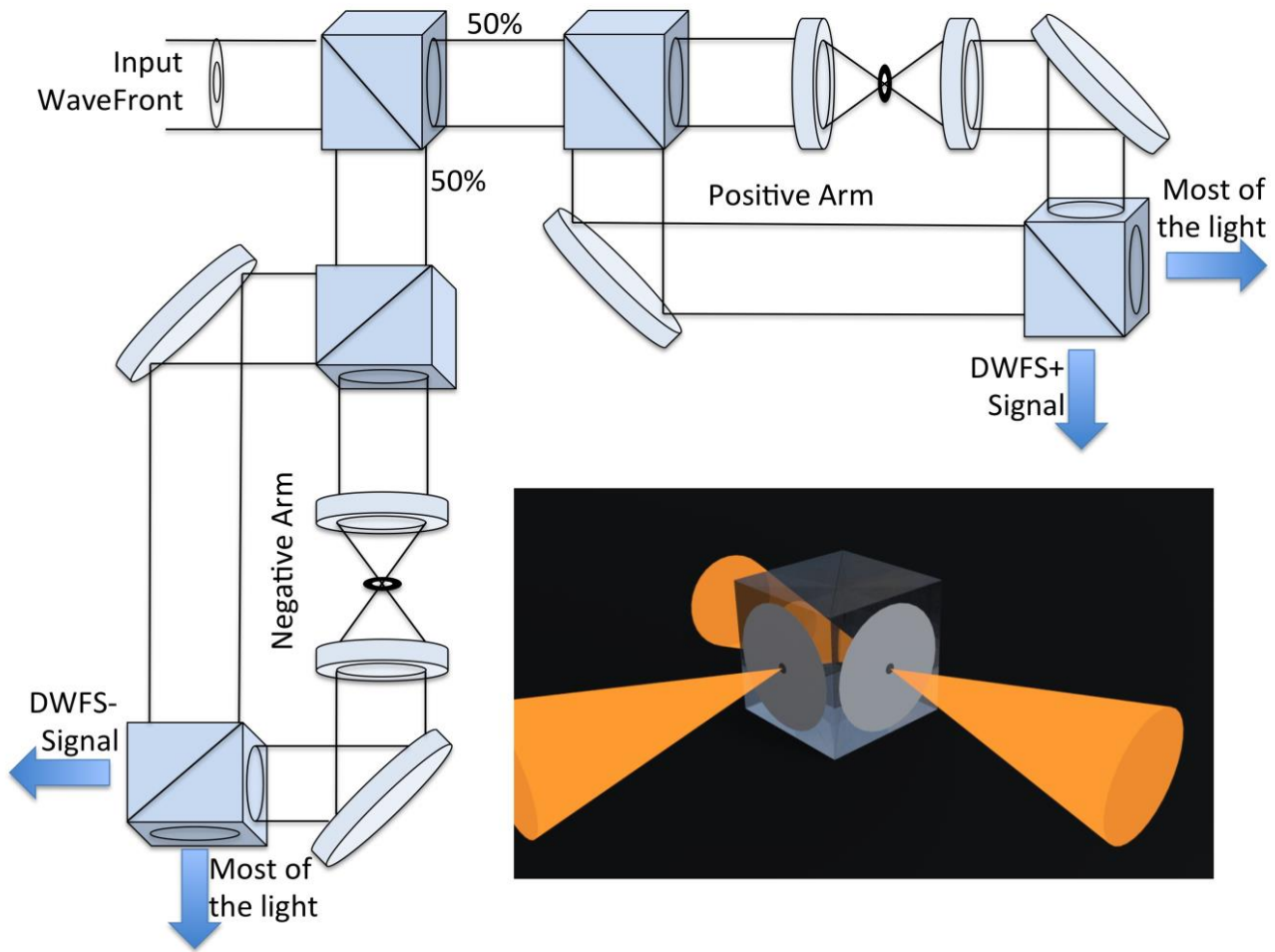

Figure 1 A sketch of an optical configuration for the DWFS. Each of the two arms generates an image of the "dark" pupil with deviation depending on the phase shift (Z-shift) applied. In the inset, the same concept is realized in a compact way with two Smartt like arms sharing in a 50/50 way the light from the reference source.

In practice the Z-shift will represent a sort of dynamic range tuner allowed for this kind of DWFS, the smaller is the smaller amount of light is getting out, but the capability to distinguish the sign of the perturbation would be more and more difficult, because of the Poissonian noise. For larger Z-shift, easier will be the distinction between a perturbation of two contradicting signs, but the WFS will not work around zero, but around a more and more significant amount of photon flux, and its corresponding Poissonian noise, something that we want to fight with this kind of device.

Of course, it not necessarily thrown away the overall amount of light, as in fact most of it, under the conditions described in the text, will go through the remaining two exit ports of the two arms, see Fig. 1. This light can be used for monitoring purposes (for instance to distinguish if by some sudden and violent tip-tilt variation the star under scrutiny simply got out from the narrow Field of View of this DWFS) or for the first order Adaptive Optics correction. The DWFS inherently works with small perturbation, so most of the light could go to a first phase "conventional" WFS (like a pyramid[7] for instance) that would keep the overall Strehl ratio high enough to guarantee that the perturbations directed into the DWFS are small enough with respect to its working range.

This matter, of how this light is treated is not being elaborated here and even in the simulations we deliberately injected, for the moment, some perturbed wavefront with characteristics that makes the DWFS working close to its optimum, by construction.

The whole optomechanical layout looks rather complex but it can be made much more compact using an optomechanical solution in which all the beams are embedded into glass, by gluing together various blocks of prisms, beam-splitters and so on, or with an even much more compact solution just by adopting on the two surfaces of a beam-splitter, two Smartt interferometers, that incorporate in a single channel the arm with the spatial filter and the unperturbed ones (see inset in Fig.1).

Of course, this would pose, further to the problem of chromaticity, that is however treated later in this paper, the one of controlling in real time the Z-shift. This, in fact, can be easily handled by some high precision (like a piezo one) positioner, in the sparse optomechanical elements solution, while in both the compact or the Smartt one could be maybe accomplished by some thermal or magneto/electrostrictive material used in the optomechanical layout. It is honest to say that while this is surely attainable with a compact, fully solid, optomechanical solution, in the Smartt case, this would act onto a very thin layer of material so it would probably pose some engineering problems or would heavily restrict the choice of the material to be used.

We do not speculate here on the further possibility that includes the spurious reflection of light, solutions that would be hard to make of high efficiency but that would multiply the capability of controlling the Z-shift in a linear manner.

.

## 3. SIMULATION CODE

The simulation code is written under the IDL environment in terms of proprietary procedures, leaving the option to incorporate this into existing packages[8][9] under such an environment as a future possibility. The incoming wavefront is treated as a bi-dimensional squared map and a corresponding mask providing the information of where actually light is being collected is associated. In the latter the information about the exact shape of the pupil, obstruction, spiders, and segmentation, if any, should be included. As usual, this would raise an issue about the finesse of such visualization, a topic that is just mention and not treated here.

The current code is monochromatic so a central wavelength is defined but it is actually the only one implemented in the current version of the code. Finally, a Z-shift is used as a free parameter, defined as a physical length. With these parameters, it is rather easy, for each of the two arms, to define the brightness of the corresponding quasi-null interferogram, sub-aperture by sub-aperture. Poissonian distribution is then applied using a completely random assignation of photon counts using as a free parameter the average flux per sub-aperture. Various tests have been performed in order to check the statistical properties of the Poissonian distribution built in the IDL language with positive success.

For the purpose of checking the correctness of the approach if no flux per sub-aperture is defined, the computed intensity (with the discretization inherent to the precision of the computation carried out) can be carried on without this Poissonian photons distribution. Checks have been done to verify that with large numbers of photons per sub-aperture the actual results are consistent with the ones obtained skipping the Poissonian distribution step.

Following the stated aims of what a DWFS should show off on the detectors, one would expect that the two maps, the "positive" and the "negative" ones would be complementary. Of course, this is –generally speaking- not true, not just because of the random deviations introduced by the statistic of the Poissonian distribution, but also because the Z-shift is a single number, that will never "fit" for all the deviations in the wavefront map. This means that often there are photons on both maps. The crude algorithm that we implemented is to select which of the two sub-aperture is brighter and to use that brightness as an estimate of the wavefront deviation. It is noticeable that in this way it is un-relevant the amount of photon collected by the "losing" sub-aperture, that would translate into the fact that we are likely not using optimally all the information collected.

No attempt has been, furthermore, made to use the unprocessed photons (for instance to flat field against scintillations effects). We just mention passing by that both these pieces of information could be used to reinforce and augment the robustness of such a DWFS, and further exploitations of these capabilities are left unexplored for the moment.

As this DWFS is intended to be used in closed loop AO operations no attempt has been made to actually calibrate the result in terms of actual deviation of the estimated wavefront, other than in the case of no Poissonian distribution of the light.

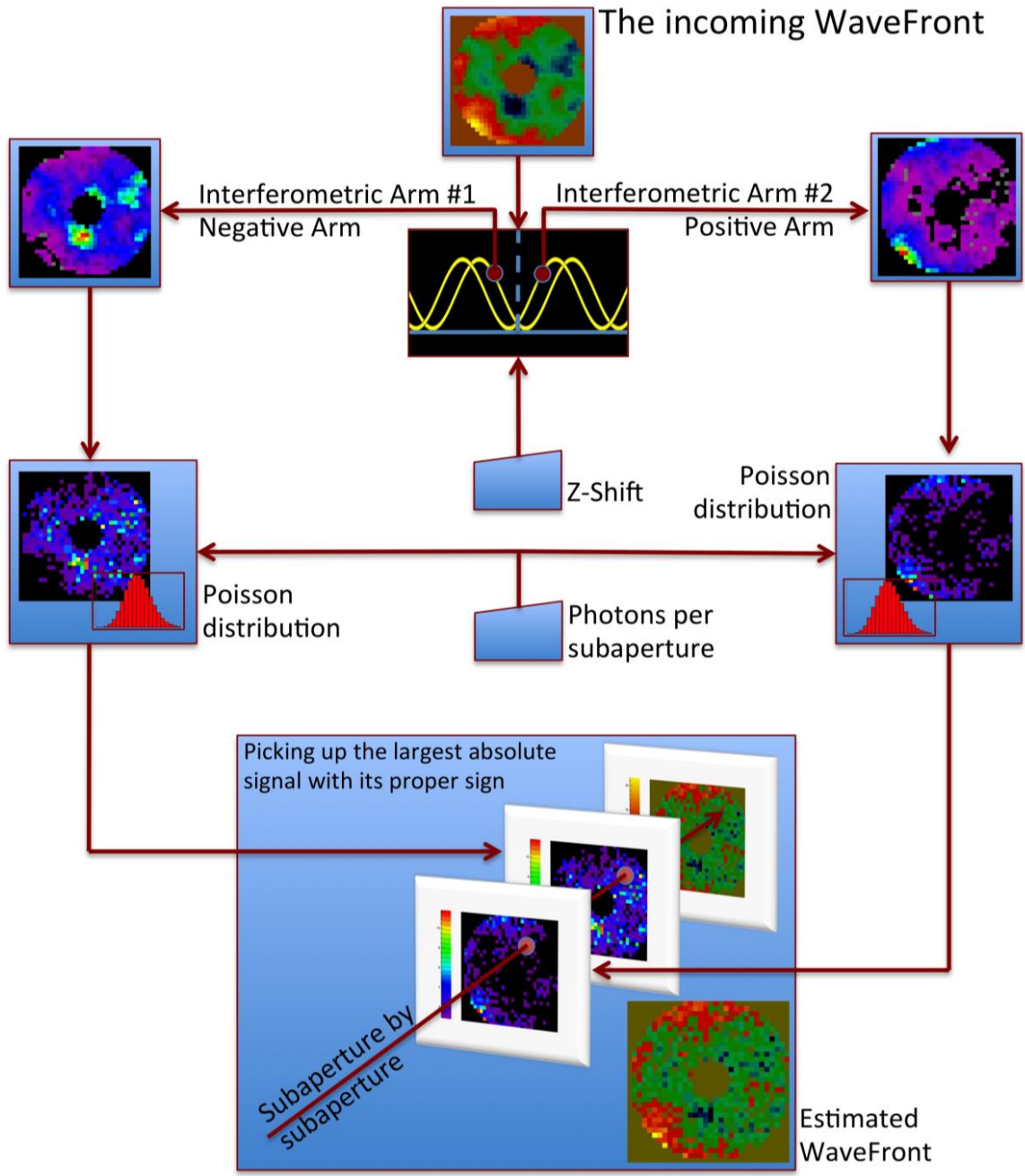

Figure 2 The flow chart of the computation algorithm used to simulate the DWFS. The incoming wavefront is used to compute the brightness distribution expected from interferometry in a quasi-nulling interferometer, with the same shift applied with two opposite signs. This leads to two identical channels of computation, named "positive" and "negative" arm. Poissonian distribution of the light is later introduced and the WaveFront is estimated with a crude approach, making the "winning" arm be the one leading to the right signal pixel per pixel.

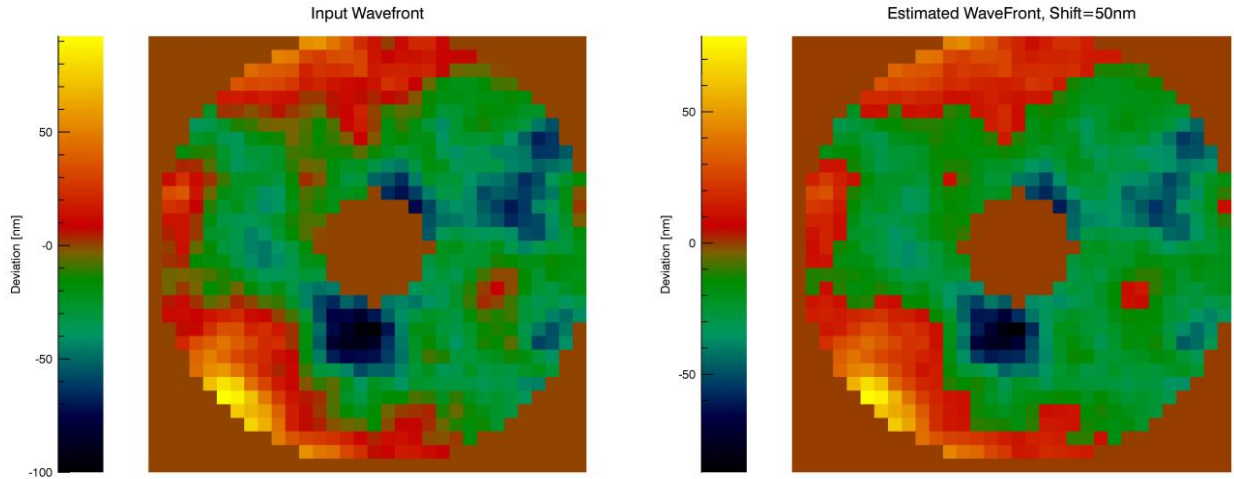

Figure 3 The input wavefront and the resulting output computed using the numerical procedure described in the text (on the left and on the right respectively). In the case shown a Z-shift of 50nm was applied.

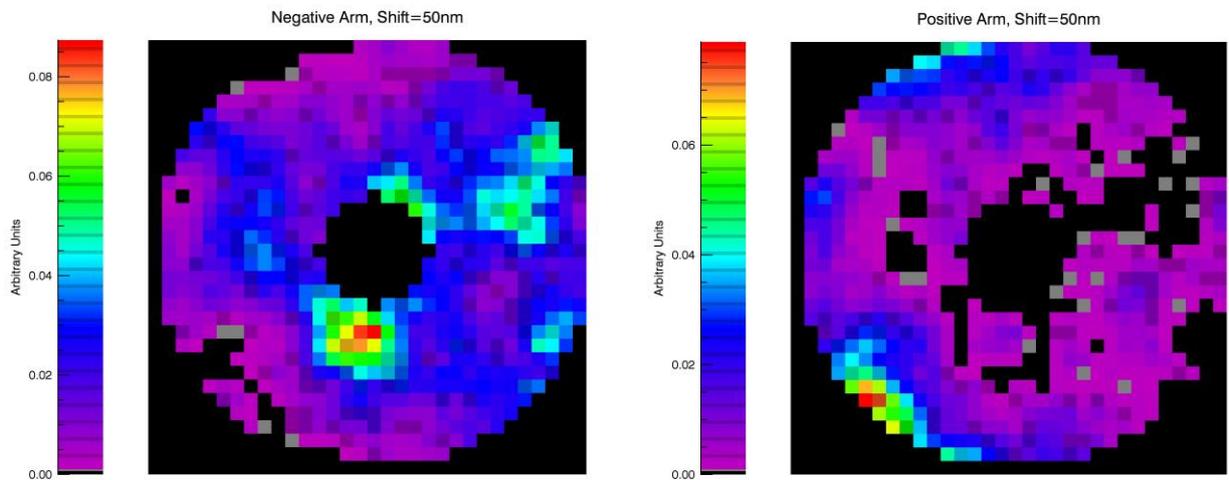

Figure 4 In the two plots above we show the resulting image at the output of the negative and positive arms. Intensity is color coded, please notice how the most delayed regions (negative part of the WF in Fig.3) show higher counts on the negative arm (on the left) and how the opposite scenario results on the positive arm.

We focused the attention to two different Z-shift values(10nm and 50nm respectively) to verify the previous statements about the effect of the Photon Noise on the measurements, see Figures from 3 to 6 and Section 2. We applied a flux of 100photon per sub-aperture.

For each of the two cases, we got the intensity map of the Negative and Positive arms. The output phase map is obtained extracting from the intensity maps the normalized values (approximation valid again for small angle/phase delays). From the comparison of the brightness intensity of the two maps, we solve the sign indetermination addressing which of the two sub-aperture of the Negative and Positive images is the winner returning the correspond sign value. This operation is done sub-aperture per sub-aperture. We actually noticed how the larger Z-shift (50nm) returns an output WF with high spatial frequency noise because of the pixel brightness error in the intensity. On the other hand, the smaller ones (10 nm) presents more spikes dues to the errors in the determination of the sign.

The proper algorithm for the sign determination is still to be refined since, for example, it is not using, any spatial information allowing the possibility to produce phase spikes: using the information on the phase sign retrieved on the close-by sub-aperture we may improve output wavefront measurements.

Moreover, we did not look into the possibility to use different Z-shift values than the ones used in the test. Another option may be the application of a slightly different Z-shift synchronized to the loop frequency.

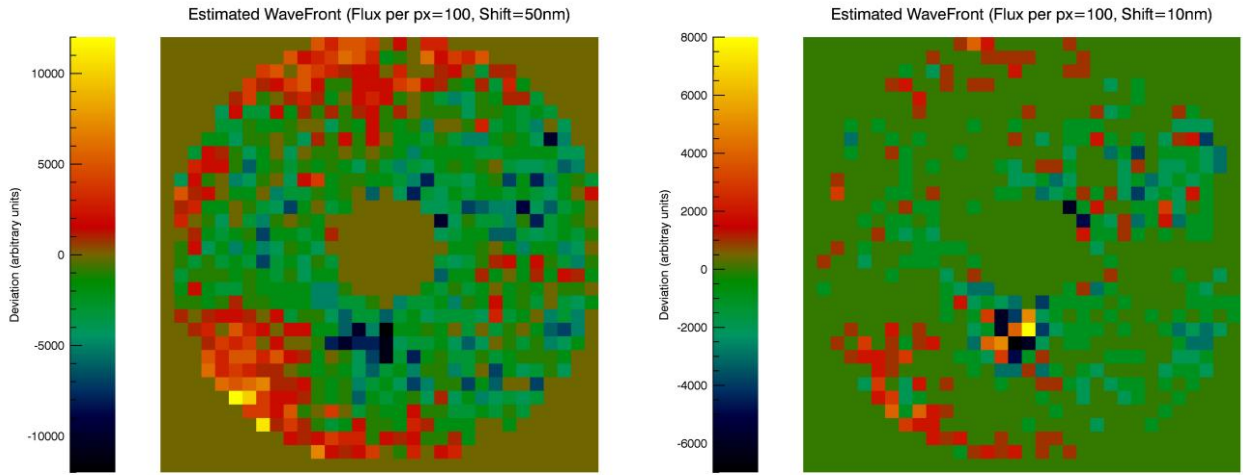

Figure 5 The pictures show two cases of WF estimation. Considering a flux of 100photon per pixel, and Z-Shift of 50nm (left) and 10nm (right). The smaller Z-Shift brought to a larger error in the determination of the sign.

## 4. FURTHER DEVELOPMENT

The selection of the Z-shift is critical in the fully exploit of the possibility of the DWFS and has to be adapted to the environment condition the DWFS has to afford. The photon flux, the power and the frequency spectrum of the input disturbance are the most important points to be considered.
In the main development plan, we foresee several steps for the upgrading of the numerical code. All upgrade points will need to develop new numerical algorithms or to implement a theoretical solution. In the following we list the possible issues regardless their actual relevance.

In the description of the numerical code we stressed that we are considering a unique and exactly monochromatic working wavelength. In order to consider the actual bandwidth we may divide the band in many different monochrome section and eventually pile up the results. In this way we may also simulate the effect of a Wynne compensator[10] opportunely inserted in the optical train that could make such a DWFS virtually a white-light one.
Nevertheless very low noise detector[11] maybe foreseen we should analyze the extra noise injected by the imaging device and with these the unavoidable effect of the sky background. We would analyze the DWFS performance within a AO closed loop system: the DWFS measurements error depends on the amplitude and spectrum of the optical disturbance and, therefore, closed-loop behavior is intimately different than open loop one. Moreover, as mentioned above, the performance of the DWFS may be enhanced by considering more Z-shift in a synchronous way respect to the read-out.
In many WFS such as pyramid an Shack-Hartmann (SH) a sort of optimization of the response may be foreseen, especially to take into account of the linearity issue (pyramid optical gain or the centroid sampling for the SH). For the DWFS may be foreseen an optimum selector of the Z-shift, either at step or continuous.
Looking into more analytic issue we would like to investigate a new implementation of the phase reconstruction algorithm that should make a better use of the information split into the negative and positive arms.

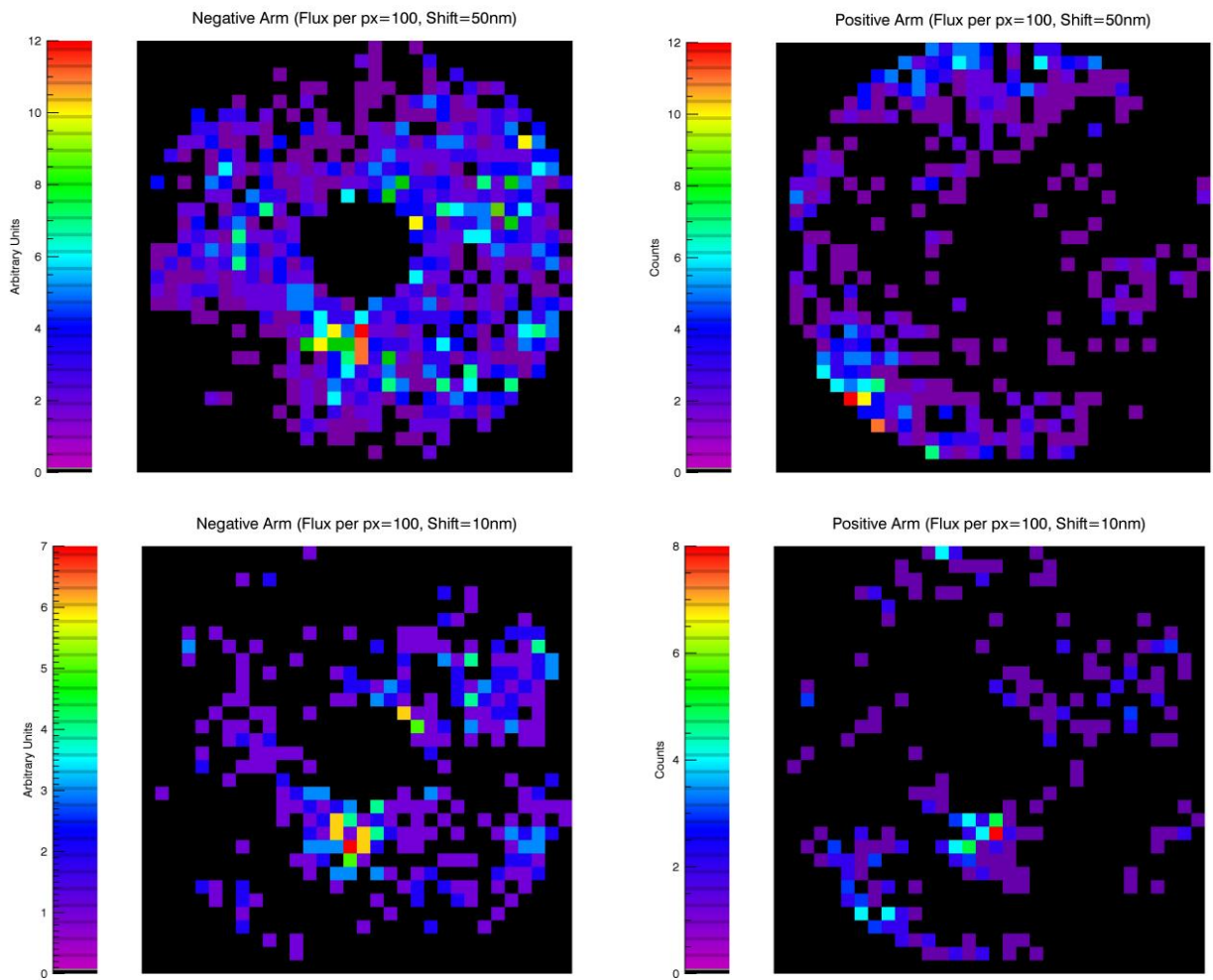

Figure 6 In the four quadrants the intensity map for the photon noise case we described. A larger Z-shift brought to a more photon populated map (50nm upper images), indeed a larger Poisson Noise too. The smaller Z-shift (bottom images) on the other side suffer less because of the Noise, but the smaller intensities variation brought to a larger sign indetermination.

## 5. CONCLUSIONS

We present here a working concept of a Dark WaveFront Sensor (DWFS). We focused on a hypothetical "perfect" realization of one of the many possible optomechanical solutions, which can realize the high photon sensitivity foreseen by the destructive interference applied to wavefront sensing. We have proven, through numerical simulations, the ability to reconstruct not only the absolute deviation of the wavefront but also its sign. We propose a double arm concept which using opposite optical path difference solves the sign indetermination looking to the intensity maps. We point out that the Z-shift (the absolute value of the optical path difference) is the key element for the optimization of the DWFS performance especially for the proper tuning with respect to flux regimes and optical disturbance amplitude.